\documentclass[conference]{IEEEtran}
\IEEEoverridecommandlockouts
\usepackage{cite}
\usepackage{geometry}
\usepackage{amsmath,amssymb,amsfonts}
\usepackage{algorithmic}
\usepackage{hyperref}
\usepackage[lined,boxed,ruled,linesnumbered]{algorithm2e}
\usepackage{graphicx}
\usepackage{textcomp}
\usepackage{xcolor}
\usepackage{bm}
\usepackage{eucal}

\newtheorem{corollary}{Corollary}
\DeclareMathOperator*{\argmax}{arg\,max}
\DeclareMathOperator*{\argmin}{arg\,min}
\begin{document}

\title{\huge{Sparse Orthogonal Matching Pursuit-based Parameter Estimation for Integrated Sensing and Communications}}
\author{%
  \IEEEauthorblockN{%
    Ngoc-Son Duong\IEEEauthorrefmark{1},
    Khac-Hoang Ngo\IEEEauthorrefmark{2},
    Thai-Mai Dinh\IEEEauthorrefmark{1}, and
    Van-Linh Nguyen\IEEEauthorrefmark{3}
  }%
  \IEEEauthorblockA{\IEEEauthorrefmark{1}Faculty of Electronics and Telecommunications, VNU University of Engineering and Technology, Hanoi, Vietnam}%
  \IEEEauthorblockA{\IEEEauthorrefmark{2}Department of Electrical Engineering (ISY), Linköping University, Linköping, Sweden}%
  \IEEEauthorblockA{\IEEEauthorrefmark{3}Department of Computer Science and Information Engineering, National Chung Cheng University, Taiwan}%
E-mails: $\lbrace${\ttfamily sondn24, dttmai}$\rbrace$@vnu.edu.vn, $\lbrace${\ttfamily khac-hoang.ngo}$\rbrace$@liu.se, $\lbrace${\ttfamily nvlinh}$\rbrace$@cs.ccu.edu.tw}
\maketitle

\begin{abstract}
Accurate parameter estimation such as angle of arrival (AOA) is essential to enhance the performance of integrated sensing and communication (ISAC) in mmWave multiple-input multiple-output (MIMO) systems. This work presents a sensing-aided communication channel estimation mechanism, where the sensing channel shares the same AOA with the uplink communication channel. First, we propose a novel orthogonal matching pursuit (OMP)-based method for coarsely estimating the AOA in a sensing channel, offering improved accuracy compared to conventional methods that rely on rotational invariance techniques. Next, we refine the coarse estimates obtained in the first step by modifying the Space-Alternating Generalized Expectation-Maximization algorithm for fine parameter estimation. Through simulations and mathematical analysis, we demonstrate that scenarios with shared AOA achieve a better Cramér-Rao lower bound (CRLB) than those without sharing. This finding highlights the potential of leveraging joint sensing and communication channels to enhance parameter estimation accuracy, particularly in channel or location estimation applications.
\end{abstract}


\section{Introduction}
Integrated sensing and communication (ISAC) will play a crucial role in the next-generation wireless networks. ISAC can enhance spectrum efficiency, reduce hardware redundancy, and improve communication performance based on sensing functions. These features are critical for enabling advanced applications like autonomous driving, extended reality, etc. Current research efforts focused on integrating sensing functionalities into existing communication systems. In particular, additional sensing parameters (extra signals) are included to increase accuracy for channel estimation \cite{b6, b7, b8, b10}. However, conventional methods often face issues such as high pilot overhead and difficulties caused by channel sparsity and mismatches.

The study in \cite{b6} is one of the first attempts to address these limitations in ISAC, significantly reducing the required training pilots by integrating spatial and temporal sensing information. The method mitigates mismatches between sensing and communication modes, a critical issue in indoor scenarios where different propagation paths or scatterers can degrade system performance. Further, the sensing-aided compressed sensing (CS) algorithm is also introduced to compensate for sensing-communication mismatches and augment missing communication modes, achieving up to a fourfold reduction in pilot overhead compared to state-of-the-art techniques. The authors in \cite{b7} proposed a sensing-aided Kalman filter (SAKF)-based channel state information (CSI) estimation method for uplink ISAC systems. This method integrates angle-of-arrival (AOA) estimation as prior information to refine least-square (LS) CSI estimation, achieving significant improvements in accuracy estimation and reduced complexity compared to traditional methods, such as minimum mean square error (MMSE) estimation. Simulation results demonstrate that the proposed method achieves bit error rates (BERs) comparable to MMSE while reducing computational overhead, showcasing its potential for efficient CSI estimation in ISAC-enabled systems. In another approach, the study in \cite{b8} introduces a novel tensor-based channel estimation method, which integrates moving target localization into the estimation process. By modeling the received signal as a high-order tensor, this method enables the decoupling and automatic pairing of channel parameters, including angles and velocities of propagation paths. High-velocity paths associated with moving targets are identified and excluded, allowing the construction of a stable communication channel using paths associated with quasi-stationary scatterers. This approach not only enhances the reliability of data transmission but also provides accurate localization of non-cooperative moving targets. Besides, the use of the compressed sensing-based framework for sensing functions was also theoretically studied in \cite{b10} and practically implemented in \cite{b2}. The authors in \cite{b10} present a CS-based framework tailored for ISAC, leveraging the inherent sparsity of \textcolor{black}{milimeter wave (mmWave)} channels. The proposed framework integrates radar sensing into communication channel estimation while reducing pilot overhead. A dual-functional radar-communication transceiver is designed with hybrid beamforming architecture, incorporating widely spaced arrays for high angular resolution in radar sensing. The study also introduces an orthogonal matching pursuit with a support refinement (OMP-SR) algorithm to mitigate angular ambiguities and ensure robust sensing and communication. Simulations demonstrate the proposed framework’s capability to improve both radar and communication performance in highly dynamic scenarios, making it a promising solution for next-generation ISAC applications.

Unlike prior work, this study presents a novel sensing-aided communication channel estimation framework in which the sensing and uplink communication channels share the same AOA. Based on the special Diriclet kernel structure of the sensing channel, we introduce a novel OMP-based algorithm that works as a coarse estimator for estimating the AOA with higher accuracy compared to the estimation of signal parameters via rotational invariant techniques (ESPRIT \cite{b5}), a well-known algorithm for AOA estimation. In the refinement step, we perform both analysis and simulation based on a modified Space-Alternating Generalized Expectation-Maximization (SAGE \cite{b12}) algorithm to show that sensing parameters can help to increase overall AOA estimation performance for both sensing and communication functions.

\textit{Notations:} Scalars are denoted by \textit{italic} letters, e.g., $x$. Lowercase boldface letters, e.g., $\textbf{x}$, denote vectors. Uppercase boldface letters, e.g., $\textbf{X}$, denote matrices. Matrix transpose, conjugate transpose, and inverse are denoted by superscripts $(\cdot)^\top$, $(\cdot)^{\text{H}}$, and $(\cdot)^{-1}$, respectively. The Euclidean norms is denoted by $\|\cdot\|_2$, respectively. Kronecker product is denoted by $\otimes$. The sets of real and complex numbers are denoted by $\mathbb{R}$ and $\mathbb{C}$, respectively. The real part of a complex number is denoted by $\mathfrak{R}\left\lbrace\cdot\right\rbrace$.
\section{System and Channel Model}
\begin{figure}[t]
    \centering
    \includegraphics[width=1\linewidth]{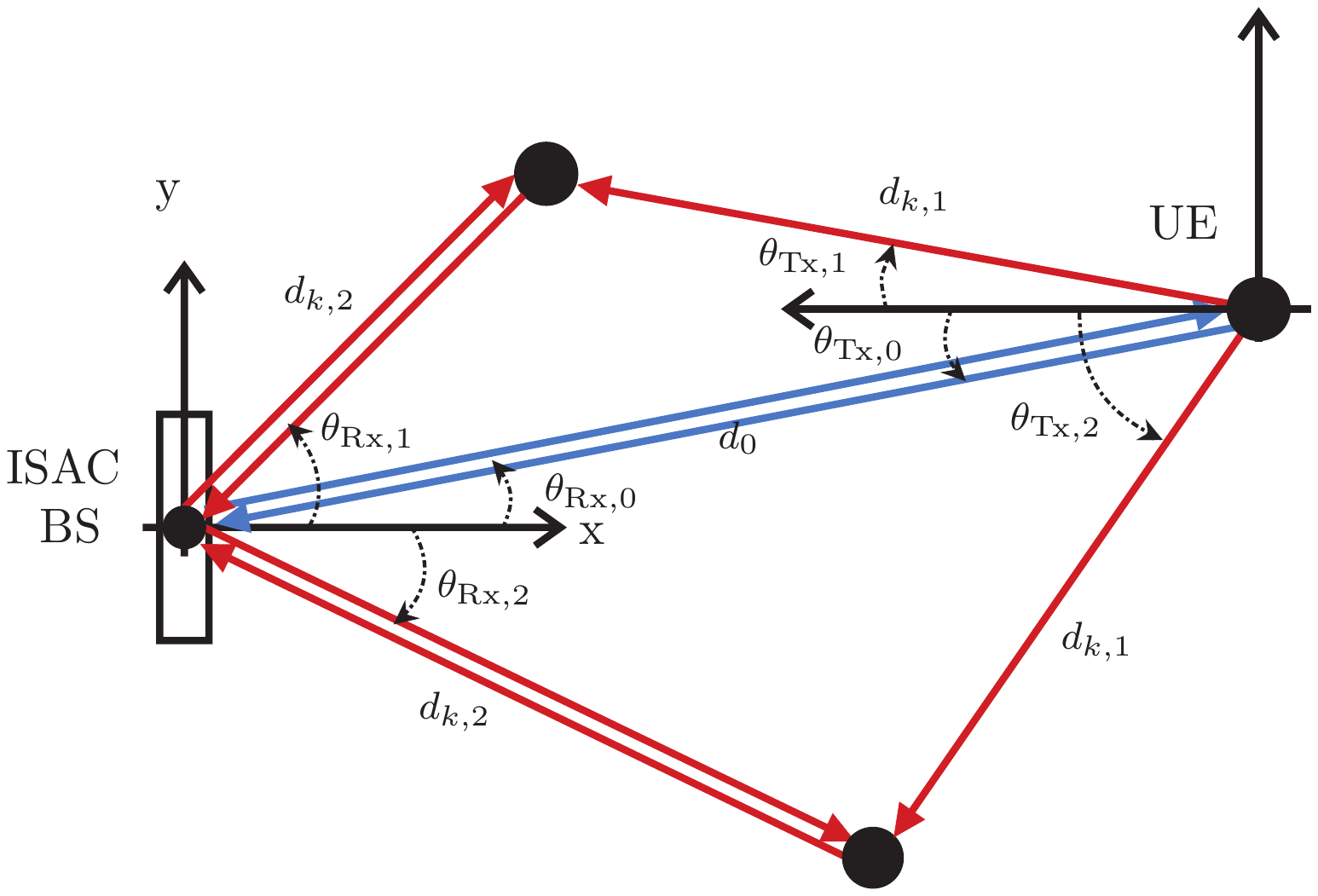}
    \caption{System model of an mmWave MIMO ISAC system.}
    \label{sys_model}
\end{figure}
This work considers an ISAC mmWave multiple-input multiple-output (MIMO) orthogonal frequency division multiplexing (OFDM) system that includes one ISAC base station (BS) and $K$ sensing targets\footnote{Although the scatter point is not always located on the sensing targets and the number of targets can be much larger, we assume the number of uplink communication paths is also $K$. As shown later in this work, such an assumption does not affect the practical channel estimation process.} and a single user equipment (UE) as shown in Fig. \ref{sys_model}.\footnote{This model can also be applied in the opposite direction, i.e., the ISAC UE tries to estimate the downlink channel while sensing the surrounding targets (a problem for intelligent vehicular systems).} In addition, we also consider the UE as a sensing target of interest. The ISAC BS employs a transmitter with $N_{t,\text{BS}}$ antennas and a collocated receiver with $N_{r,\text{BS}}$ antennas, where the antennas are arranged in uniform linear arrays (ULA) with half-wavelength intervals. Similarly, the UE antenna is assumed to be a ULA with $N_{t,\text{UE}}$ elements. For simplicity, we write $N_{t,\text{UE}} = N_t$ and assume $N_{t,\text{BS}} = N_{r,\text{BS}} = N_r$. The steering and response vectors of the ULA are generally given by
\begin{equation}
    \textbf{a}(\theta) = \left[ 1, e^{-j\pi\sin(\theta)},...,e^{-j\pi(M-1)\sin(\theta)}\right]^\top,
\end{equation}
where $\theta$ and $M$ are the angle parameter and the number of antenna elements. In the considered ISAC system, the transmitter emits sensing signals for environment sensing and the UE transmits uplink pilots for channel estimation. Meanwhile, the receiver simultaneously collects the echo signals and the uplink pilots. At first, the BS estimates sensing and uplink communication channels separately through typical sparse methods, e.g., OMP, Sparse Bayesian Learning (SBL), Approximate Message Passing (AMP), etc. Herein, we assume that self-interference is canceled out and beam squint effects are negligible. The results in this step are only coarse values because the angle parameters are estimated on discrete bases. After that, the BS performs a refinement step to obtain off-grid parameters for both subsystems. In addition, we assume the channel remains constant during estimation time.

\subsection{Sensing Channel}
For the sensing part, the collected echo signals by the BS can be expressed as
\begin{equation}
    \mathbf{y}_s[n] = \mathbf{H}_s[n]\mathbf{s}[n] + \mathbf{n}_s[n],
\end{equation}
where $n \in [N]$ indicates the subcarrier index while $\mathbf{H}_s$, $\mathbf{s}$ and $\textbf{n}_s \sim \CMcal{CN}(0, \sigma_s^2\textbf{I})$ are sensing channel, sensing pilot, and the additive noise, respectively. Herein, the sensing channel matrix is expressed as
\begin{equation}
    \mathbf{H}_s = \sum_{k = 1}^{K}{\alpha_k
    \mathbf{a}^{\text{H}}(\theta_{\text{Rx},k})\mathbf{a}(\theta_{\text{Tx},k})},
\end{equation}
where $\alpha_k$ denotes channel gain of the $k$-th sensing path. For each target, we assume AOA, denoted by $\theta_{\text{Rx},k}$, is the same as angle of departure (AOD), denoted by $\theta_{\text{Tx},k}$. Moreover, the model does not include any beamforming or combining matrix.

\subsection{Communication Channel}
For the communication part, the signal transmitted over the $n$-th subcarrier can be written as $\textbf{F}[n]\textbf{x}[n]$, where $\textbf{F}[n]$ and $\textbf{x}[n]$ are the beamforming matrix and known pilot signal, respectively. With respect to the $n$-th subcarrier, the channel matrix is written as
\begin{equation}
    \mathbf{H}_c = \sum_{k = 1}^{K}{\beta_k
    \mathbf{a}^{\text{H}}(\varphi_{\text{Rx},k})\mathbf{a}(\varphi_{\text{Tx},k})},
\end{equation}
where $\varphi_{\text{Rx},k}$ and $\varphi_{\text{Tx},k}$ are AOA and AOD of $k$-th uplink communication path, respectively. Meanwhile, $\beta_k$ is the complex gain of the $k$-th communication path. The received signal can be written as
\begin{equation}
\textbf{y}_c[n] = \textbf{H}_c[n]\textbf{F}[n]\textbf{x}[n] + \textbf{n}_c[n],
\label{eq7}
\end{equation}
where $\textbf{n}_c[n] \sim \CMcal{CN}(0, \sigma_c^2\textbf{I})$ is the additive noise.

\section{Proposed sensing-aided parameter estimation for mmWave MIMO ISAC system}

We initially transfer the mmWave MIMO channel to the angular domain in order to expose its sparsity. We divide the domain $\left[ -\frac{\pi}{2}, \frac{\pi}{2}\right)$ into a grid with resolution $G_t \geq N_t$ and $G_r \geq N_r$, where $i$-th and $j$-th grid point is denoted by $\overline{\omega}_i$ and $\overline{\omega}_j$, respectively. We then introduce two unitary transformation matrices $\textbf{U}_{\text{Tx}}$ and $\textbf{U}_{\text{Rx}}$ as
\begin{equation}
\textbf{U}_{\text{Tx}} \triangleq [\textbf{u}_{\text{Tx}}(\overline{\omega}_1), \textbf{u}_{\text{Tx}}(\overline{\omega}_2),..., \textbf{u}_{\text{Tx}}(\overline{\omega}_{G_t})],
\label{eq8}
\end{equation}
\begin{equation}
\textbf{U}_{\text{Rx}} \triangleq [\textbf{u}_{\text{Rx}}(\overline{\omega}_1), \textbf{u}_{\text{Rx}}(\overline{\omega}_2),..., \textbf{u}_{\text{Rx}}(\overline{\omega}_{G_r})],
\label{eq9}
\end{equation}
where $\textbf{u}_{\text{Tx}}(\overline{\omega}_i)$, $\textbf{u}_{\text{Rx}}(\overline{\omega}_i)$ are called the atoms of $\textbf{U}_{\text{Tx}}$ and $\textbf{U}_{\text{Rx}}$, respectively. After multiply $\textbf{U}_{\text{Tx}}$ and $\textbf{U}_{\text{Rx}}$ to the communication channel, we obtain
\begin{equation}
\textbf{y}_c[n] = \bm{\Omega}_c[n]\textbf{h}_c[n] + \textbf{n}_c[n],
\end{equation}
where $\bm{\Omega}_c[n] = \big( \textbf{U}_{\text{Tx}}^{\text{H}}\textbf{F}[n]\textbf{x}[n]\big)^\top \otimes \textbf{U}_{\text{Rx}}$ and $\textbf{h}_c[n] = \text{vec}(\textbf{U}_{\text{Rx}}^{\text{H}}\textbf{H}_c[n]\textbf{U}_{\text{Tx}})$. With $\mathbf{h}_c$ being a sparse vector, we need to find
\begin{align}
\textbf{h}_c[n] = &\argmin_{{\mathbf{h}_c}[n]} ~~\Vert{\textbf{h}_c}[n]\Vert_0\label{sparse_comm} \\
&\text{subject to} \quad \Vert {\textbf{y}_c}[n] - \bm{\Omega}_c[n]{\textbf{h}_c}[n]\Vert \leq \epsilon_c,
\end{align}
where $\epsilon_c$ is a small positive constraint. Similarly, we obtain an equivalent problem for the sensing function by replacing subscript $c$ by $s$. For the sensing problem, we have $\bm{\Omega}_s[n] = \big( \textbf{U}_{\text{Tx}}^{\text{H}}\textbf{s}[n]\big)^\top \otimes \textbf{U}_{\text{Rx}}$ and $\textbf{h}_s[n] = \text{vec}(\textbf{U}_{\text{Rx}}^{\text{H}}\textbf{H}_s[n]\textbf{U}_{\text{Tx}})$.

\subsection{Proposed OMP Algorithm for Sensing Function}

The orthogonal matching pursuit (OMP) algorithm is a widely used greedy approach for sparse signal reconstruction in compressed sensing. It can find optimal solutions for \eqref{sparse_comm} as long as $\bm{\Omega}_c[n]$ and $\bm{\Omega}_s[n]$ satisfy a so-called restricted isometry property \cite{b11}. OMP iteratively selects dictionary atoms that best correlate with the residual signal and refines the solution by solving a least-squares problem. This process continues until a predefined sparsity level or error threshold is met, offering a balance between computational efficiency and reconstruction accuracy. When applied to our problem, OMP is modified and improved in the following way. It starts with an atom selection step based on the correlation of all atoms with the residual signal as
\begin{equation}
    \left[m^{\text{max}}, c\right] = \argmax_m \sum_{n = 1}^{N}{\vert\langle\textbf{r}[n], \mathbf{\Omega}_m[n]\rangle\vert},
    \label{eq:corr}
\end{equation}
where \textcolor{black}{$\textbf{r}[n]$ and $\mathbf{\Omega}_m[n]$ are the residual and $m$-th column of sensing matrix on $n$-th subcarrier, respectively. In the left hand side,} $m^{\text{max}}$ is the position at which the correlation reaches its highest value, denoted by $c$. However, since AOA and AOD are the same in the sensing part, the possible set for the OMP-based sensing algorithm reduces to
\begin{equation}
    \CMcal{J} = \lbrace m \vert m = i + G_r(j-1) \land i = j\rbrace,
\end{equation}
where $i, j$ are AOA and AOD indices, which correspond to quantized AOA and AOD in the beamspace domain, respectively. Generally, the OMP algorithm will estimate the channel coefficient for an on-the-fly atom. However, with the limited number of antennas, the channel in the beamspace domain is not completely sparse \cite{b4} and hence leads to a phenomenon called power leakage \cite{b10}. Specifically, the amplitude spectrum of each path spans multiple grids, preventing a single atom from capturing the entire power of a given path. Consequently, the residual from the remaining paths still contains some power from the previously identified path, leading to the failure of OMP in accurately recovering the full support set. Instead of reacting to power leakage as a phenomenon to be avoided \cite{b10}, we reconcile with it. The use of redundant dictionaries can be considered counterproductive, as it increases algorithmic complexity without improving estimation accuracy. This is due to the increased number of grid points, which raises the mutual coherence \cite{b11} and may even degrade the performance of the sparse recovery algorithm. In this case, we exploit the structure of the sparse sensing channel. Specifically, in the beamspace domain, the channel exhibits a two-dimensional Dirichlet kernel structure \cite{b4}, where each path is characterized by a main lobe accompanied by multiple side lobes. To address this situation, we utilize nearby supports within the main lobe to estimate their channel coefficients. In each iteration, the on-the-fly atom is treated as the central position, and additional selected atoms are constrained to lie within a specified radius
\begin{equation}
    a = \bigg\lfloor \frac{G_r}{4\pi}\bigg\rfloor.
\end{equation}
Then the set of selected atoms in one loop is
\begin{equation}
    \CMcal{J}' \subset \CMcal{J} = \lbrace m \vert m^{\text{max}} - aG_r - a,...,m^{\text{max}},..., m^{\text{max}} + aG_r + a  \rbrace.\label{subset}
\end{equation}
Once the supports are selected, we evaluate their contribution to the power of a single path. Here, we choose the correlation values in set $\CMcal{J}'$ as the weight values, i.e., the higher the correlation value, the greater the contribution of that atom and vice versa. The weight corresponding to the $m$-th atom is given by
\begin{equation}
    w(m) = \frac{c(m)}{\sum_m{c(m)}}, \forall m \in \CMcal{J}'.
\end{equation}
After that, we update the support set as $\CMcal{I} = \CMcal{I} \cup \CMcal{J}'$ and estimate the channel as
\begin{equation}
    \hat{\mathbf{h}}_s[n] = \left(\text{diag}(\mathbf{w})\mathbf{\Omega}_{\CMcal{I}}^\text{H}[n] \mathbf{\Omega}_{\CMcal{I}}[n]\right)^{-1} \text{diag}(\mathbf{w})\mathbf{\Omega}_{\CMcal{I}}^\text{H}[n] \mathbf{y}_s[n],
\end{equation}
where $\mathbf{w} \in \mathbb{R}^{2a + 1}$ is the weight vector that includes element $w(m)$. Finally, we update the residual as
\begin{equation}
    \mathbf{r}[n] = \mathbf{y}_s[n] - \mathbf{\Omega}_{\CMcal{I}}[n] \hat{\mathbf{h}}_s[n].
\end{equation}
The proposed method is summarized in Algorithm 1. The main difference between the proposed OMP algorithm and the conventional OMP comes from the structure of the channel in the beamspace domain. When considering a non-zero position, the proposed OMP algorithm considers not only the position with the highest correlation value but also the values around it (Line 4 and 6 in Algorithm 1).
\begin{algorithm}[t]
\SetAlgoLined
	\caption{Proposed OMP for sensing channel estimation}\label{ttomp}
    \SetKwInOut{Input}{Input}
    \SetKwInOut{Output}{Output}
    \Input{Echo signal: $\textbf{y}_s$, Sensing matrix: $\mathbf{\Omega} = \mathbf{\Omega}_s$ \\ Sparsity: $K$, Tolerance: $\epsilon$ and Beamwidth: $a$}
    \SetKwInOut{Initialize}{Initialize}
    \Initialize{Support set $\CMcal{I}_0 = \left\lbrace\emptyset\right\rbrace$, residual $\mathbf{r}_0 = \mathbf{y}_s$}
    \While{$\Vert\mathbf{r}^{(t)}\Vert_2 \geq \epsilon$ \text{or} $t \leq K $}{
    1. Choosing atom $\left[m^{\text{max}}, c\right] = \argmax_m \sum_{n = 1}^{N}{\vert\langle\textbf{r}^{(t)}[n], \mathbf{\Omega}_m[n]\rangle\vert}$;\\
    2. Generating $\CMcal{J}' = \lbrace m \vert m^{\text{max}} - aG_r - a,...,m^{\text{max}},..., m^{\text{max}} + aG_r + a \rbrace$;\\
    3. Weighting $\mathbf{w} \leftarrow w(m) = \frac{c(m)}{\sum_j{c(m)}}, \forall m \in \CMcal{J}'$;\\
    4. Updating support set $\CMcal{I}^{(t)} = \CMcal{I}^{(t-1)} \cup \CMcal{J}'$;\\
    5. Estimating current channel $\hat{\mathbf{h}}_s^{(t)}[n] = \left(\text{diag}(\mathbf{w})\mathbf{\Omega}_{\CMcal{I}^{(t)}}^\text{H}[n] \mathbf{\Omega}_{\CMcal{I}^{(t)}}[n]\right)^{-1} \text{diag}(\mathbf{w})\mathbf{\Omega}_{\CMcal{I}^{(t)}}^\text{H}[n] \mathbf{y}_s[n]$;\\
    6. Update residual $\mathbf{r}^{(t)}[n] = \mathbf{y}_s[n] - \mathbf{\Omega}_{\CMcal{I}^{(t)}}[n] \mathbf{h}^{(t)}[n]$;\\
    }
    \Output{$\CMcal{I}, \hat{\mathbf{h}}_s[n]$}
\end{algorithm}

\textbf{\textit{Remark}}: Although the Dirichlet kernel structure includes side lobes, we only use the main lobe as an approximation for two reasons: \textit{i)} the atoms of side lobes may belong to another main lobe and sometimes the correlation values corresponding to those atoms are not exact because of the superposition and \textit{ii)} this reduces the complexity of the algorithm.
\subsection{Proposed strategy for refinement procedure}
Let $\bm{\eta} = \left\lbrace \bm{\alpha},\bm{\beta}, \bm{\varphi}_{\text{Tx}}, \bm{\varphi}_{\text{Rx}}, \bm{\theta}_{\text{Tx}}, \bm{\theta}_{\text{Rx}}\right\rbrace$ be a set of all parameter, refinement process is performed simultaneously for both sensing and communication function when the sparse estimation algorithms find that they have the same AOA. Assuming $N = 1$ for simplicity, the LS cost function for the joint case is given by
\begin{equation}
    \CMcal{Q}_{\text{joint}} = \Vert\mathbf{y}_c + \mathbf{y}_s - \mathbf{\Omega}_c\mathbf{h}_c - \mathbf{\Omega}_s\mathbf{h}_s\Vert_2^2.\label{qj}
\end{equation}
Otherwise, it is estimated separately with two different cost functions as
\begin{equation}
    \CMcal{Q}_{c} = \Vert\mathbf{y}_c - \mathbf{\Omega}_c\mathbf{h}_c\Vert_2^2\label{qc}
\end{equation}
and
\begin{equation}
    \CMcal{Q}_{s} = \Vert\mathbf{y}_s - \mathbf{\Omega}_s\mathbf{h}_s\Vert_2^2\label{qs}
\end{equation}
To minimize \eqref{qj}, \eqref{qc}, and \eqref{qs}, we use a modified SAGE algorithm as shown in Algorithm 2. In the E-step (Line 3 in Algorithm 2), it re-calibrates the cost function using the found parameters of $K$ paths in the previous loop and considers the remaining one as an unknown parameter. The algorithm next finds an optimal solution for the cost function using gradient descent in the M-step (Line 6 in Algorithm 2). In this work, when optimizing AOA (including $\bm{\varphi}_{\text{Rx}}, \bm{\theta}_{\text{Tx}}, \bm{\theta}_{\text{Rx}}$), the update rule is
\begin{equation}
    \hat{\eta}_{p,k}^{(i)} = \hat{\eta}_{p,k}^{(i-1)} - \gamma_{\eta_{p,k}}\frac{\partial \CMcal{Q}_{\text{joint}}}{\partial \eta_{p,k}}.
\end{equation}
Herein, $\gamma_{\eta_{p,k}}$ is a descent coefficient, $\frac{\partial \CMcal{Q}_{\text{joint}}}{\partial \eta_{p,k}}$ is calculated as in \eqref{daoham}, where $\mathbf{z}_{c,l} = \mathbf{y}_c - \sum_{l=1,l\neq k}^{K}{\mathbf{\Omega}_c\mathbf{h}_{c,l}}$ and $\mathbf{z}_{s,l} = \mathbf{y}_s - \sum_{l=1,l\neq k}^{K}{\mathbf{\Omega}_s\mathbf{h}_{s,l}}$.
\begin{figure*}[ht]
\begin{equation}
    \frac{\partial \CMcal{Q}_{\text{joint}}}{\partial \eta_{p,k}} = -2 \mathfrak{R} \left\lbrace \left( \mathbf{\Omega}_c^{\text{H}} (\mathbf{z}_{c,l} - \mathbf{\Omega}_c\mathbf{h}_{c,k} + \mathbf{z}_{s,l} - \mathbf{\Omega}_s\mathbf{h}_{s,k}) \right)^{\text{H}} \frac{\partial \mathbf{h}_c}{\partial \eta_{p,k}} \right\rbrace 
    -2 \mathfrak{R} \left\lbrace \left( \mathbf{\Omega}_s^{\text{H}} (\mathbf{z}_{c,l} - \mathbf{\Omega}_c\mathbf{h}_{c,k} + \mathbf{z}_{s,l} - \mathbf{\Omega}_s\mathbf{h}_{s,k}) \right)^{\text{H}} \frac{\partial \mathbf{h}_s}{\partial \eta_{p,k}} \right\rbrace\label{daoham}
\end{equation}
\hrule
\end{figure*}
In contrast, optimizing other parameters involves only one of the two cost functions of individual systems, i.e, \eqref{qc} or \eqref{qs}. The update rule for parameters that only belong to the communication part is
\begin{equation}
    \hat{\eta}_{p,k}^{(i)} = \hat{\eta}_{p,k}^{(i-1)} - \gamma_{\eta_{p,k}}\frac{\partial \CMcal{Q}_c} {\partial \eta_{p,k}},
\end{equation}
where, $\frac{\partial \CMcal{Q}_c}{\partial \eta_{p,k}}$ is calculated similarly as \eqref{daoham} by setting $\mathbf{z}_{s,l}, \mathbf{h}_{s,k}$, and $\mathbf{h}_s$ to zero. Likewise, we can obtain the update rule for parameters that only belong to the sensing part by replacing subscript $c$ by $s$. Meanwhile, the sensing channel parameters are initialized by the proposed OMP algorithm, while the communication channel parameters are initialized by the Distributed Compressed Sensing Simultaneous OMP (DCS-SOMP) algorithm \cite{b1}.
\begin{algorithm}[t]
\SetAlgoLined
	\caption{SAGE \cite{b12} as Coordinate Descent plus Gradient Descent algorithm for the refinement step (AOA-shared case with $N = 1$)}\label{sage}
    \SetKwInOut{Input}{Input}
    \Input{$\textbf{y}_{c}$, $\textbf{y}_{s}$, $\mathbf{\Omega}_c$, $\mathbf{\Omega}_s$, $M$}
    \SetKwInOut{Initialize}{Initialize}
    \Initialize{$\bm{\eta}^{(0)}$, $\textbf{y}_{c}^{(0)}$, $\textbf{y}_{s}^{(0)}$, $\mathbf{h}_c^{(0)}$, $\mathbf{h}_s^{(0)}$}
    \For{$i = 1,...,M$}{
    \For{$k = 1,..., K$}{
        \textit{E-step:} \\ 
        $\mathbf{z}_{c,l}^{(i)} = \mathbf{y}_c - \sum_{l=1,l\neq k}^{K}{\mathbf{\Omega}_c\mathbf{h}_{c,l}^{(i-1)}}$;\\
        $\mathbf{z}_{s,l}^{(i)} = \mathbf{y}_s - \sum_{l=1,l\neq k}^{K}{\mathbf{\Omega}_s\mathbf{h}_{s,l}^{(i-1)}}$;\\
        \textit{M-step:} \\
    \For{$p = 1,...,6$}{
		$\hat{\eta}_{p,k}^{(i)} = \hat{\eta}_{p,k}^{(i-1)} - \gamma_{\eta_{p,k}}\frac{\partial \CMcal{Q}_{\text{joint}}}{\partial \eta_{p,k}}$ if $\eta_{p,k}$ are $\bm{\varphi}_{\text{Rx}}, \bm{\theta}_{\text{Tx}}$, and $\bm{\theta}_{\text{Rx}}$; \\

        $\hat{\eta}_{p,k}^{(i)} = \hat{\eta}_{p,k}^{(i-1)} - \gamma_{\eta_{p,k}}\frac{\partial \CMcal{Q}_c}{\partial \eta_{p,k}}$ if $\eta_{p,k}$ are $\bm{\alpha}$ and $\bm{\varphi}_{\text{Tx}}$;

        $\hat{\eta}_{p,k}^{(i)} = \hat{\eta}_{p,k}^{(i-1)} - \gamma_{\eta_{p,k}}\frac{\partial \CMcal{Q}_s}{\partial \eta_{p,k}}$ if $\eta_{p,k}$ is $\bm{\beta}$;
	   }
    }
    Reconstruct $\mathbf{h}_c^{(i)}$ and $\mathbf{h}_s^{(i)}$;\\
}
    \SetKwInOut{Output}{Output}
    \Output{$\bm{\eta}^{(M)}$}
\end{algorithm}
Our problem can be seen as a cooperative problem in which many systems share some common parameters. In this case, the Cramér-Rao Lower Bound (CRLB) of the shared parameter satisfies a property in \cite{b13}, where the CRLB of a parameter decreases when the number of measurements increases. It is tailored for our problem as the following corollary.
\begin{corollary}
\label{th1}
\textit{Considering a system consisting of two subsystems, for example, an uplink communication system $S_1$ and a sensing system $S_2$. If the subsystems share a common parameter $\bm{\theta}$, then the CRLB of shared $\bm{\theta}$ is lower than elemental CRLBs of $\bm{\theta}$ assuming they operate independently, sharing no parameter, i.e.}
\begin{equation}
    \text{CRLB}_{\text{shared}}(\bm{\theta}) \leq \min \left\lbrace\text{CRLB}_{S_1}(\bm{\theta}), \text{CRLB}_{S_2}(\bm{\theta})\right\rbrace.
\end{equation}
\end{corollary}
\textit{Proof:} The joint likelihood for $\mathbf{y}_{c}$ and $\mathbf{y}_{s}$ in general case is
\begin{equation}
    \CMcal{L}(\mathbf{y}_c, \mathbf{y}_s \mid \bm{\theta}) = \CMcal{L}(\mathbf{y}_c \mid \mathbf{y}_s,\bm{\theta}) \CMcal{L}(\mathbf{y}_s \mid \bm{\theta})
\end{equation}
Then, the joint log-likelihood is
\begin{equation}
    \ln \CMcal{L}(\mathbf{y}_c, \mathbf{y}_s \mid \bm{\theta}) = \ln \CMcal{L}(\mathbf{y}_c \mid \mathbf{y}_s, \bm{\theta}) + \ln \CMcal{L}(\mathbf{y}_s \mid \bm{\theta}),
\end{equation}
The Fisher information (FI) for $\bm{\theta}$ is the sum of the FI contributions from both systems
\begin{equation}
    \CMcal{I}(\bm{\theta}) = \CMcal{I}_1(\bm{\theta}) + \CMcal{I}_2(\bm{\theta}),
\end{equation}
where
\begin{equation}
    \CMcal{I}_1(\bm{\theta}) = -\sum_{k=1}^{K}{\mathbb{E}\left[\frac{\partial^2 \ln \CMcal{L}(\mathbf{y}_c \mid  \mathbf{y}_s, \theta_k)}{\partial \theta_{k}^2}\right]} > 0,
\end{equation}
and
\begin{equation}
    \CMcal{I}_2(\bm{\theta}) = -\sum_{k=1}^{K}{\mathbb{E}\left[\frac{\partial^2 \ln \CMcal{L}_2(\mathbf{y}_s \mid \theta_{k})}{\partial \theta_{k}^2}\right]} > 0.
\end{equation}
The CRLB for the AOA parameter $\bm{\theta}$ in the joint case is
\begin{equation}
    \text{CRLB}(\hat{\bm{\theta}}) = \frac{1}{\CMcal{I}(\bm{\theta})} = \frac{1}{\CMcal{I}_1(\bm{\theta}) + \CMcal{I}_2(\bm{\theta})}.
\end{equation}
Meanwhile, if we treat $\bm{\theta}$ in two systems as separated parameters, the CRLB of AOA of each separated function are $\text{CRLB}(\hat{\bm{\theta}}_{S_1}) = \frac{1}{\CMcal{I}_1(\bm{\theta})}$, and $\text{CRLB}(\hat{\bm{\theta}}_{S_2}) = \frac{1}{\CMcal{I}_2(\bm{\theta})}$. Finally, we obtain straightforwardly that
\begin{equation}
    \text{CRLB}(\hat{\bm{\theta}}_{\text{shared}}) = \frac{1}{\CMcal{I}_1(\bm{\theta}) + \CMcal{I}_2(\bm{\theta})} < \frac{1}{\CMcal{I}_2(\bm{\theta})} = \text{CRLB}(\hat{\bm{\theta}}_{S_2}).
\end{equation}
Similarly, interchanging $\mathbf{y}_c$ and $\mathbf{y}_s$, we obtain
\begin{equation}
    \text{CRLB}(\hat{\bm{\theta}}_{\text{shared}}) <  \text{CRLB}(\hat{\bm{\theta}}_{S_1}).
\end{equation}
\section{Simulation results and discussion}
The number of transmit and receive antennas are set to $N_t = N_r = 16$ elements. The number of virtual beams is $G_t = G_r = 32$. The channel includes one LOS and one NLOS path. We employ successful recovery probability (SRP) and root mean square error (RMSE) as two indicators to assess the estimation performance as
\begin{equation}
\text{SRP} = \frac{\text{number of successful trial}}{\text{total number of trial}},
\end{equation}
where a trial is considered successful if the algorithm detects exactly the support set, and
\begin{equation}
    \text{RMSE}(\theta) = \sqrt{\mathbb{E}(\hat{\theta} - \theta)}.
\end{equation}
  The performance is evaluated under varying SNR conditions. The SRP and RMSE of the estimation algorithms are averaged over 1000 Monte Carlo realizations.
\subsection{Coarse Estimation Performance}
\begin{figure}[t]
    \centering
    \includegraphics[width=1.05\linewidth]{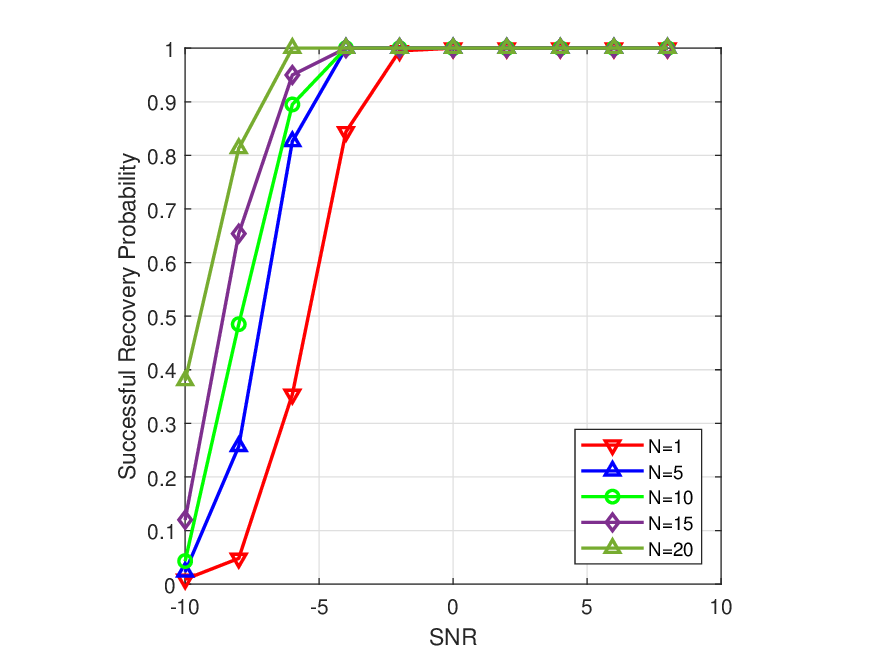}
    \caption{Successful recovery probability versus SNR (dB) in the coarse estimation step}
    \label{rs1}
\end{figure}
The SRP performance of the proposed method are shown in Fig. \ref{rs1}. The key observation from this result is that the SRP improves as the number of subcarriers increases. As shown in Fig. \ref{rs1}, the SRP go to 1 as the SNR becomes large since the noise present in $\textbf{r}$ is white noise. Increasing $N$ progressively reduces the noise power in \eqref{eq:corr}, leading to an increase in SRP. 
\begin{figure}[t]
    \centering
    \includegraphics[width=1.05\linewidth]{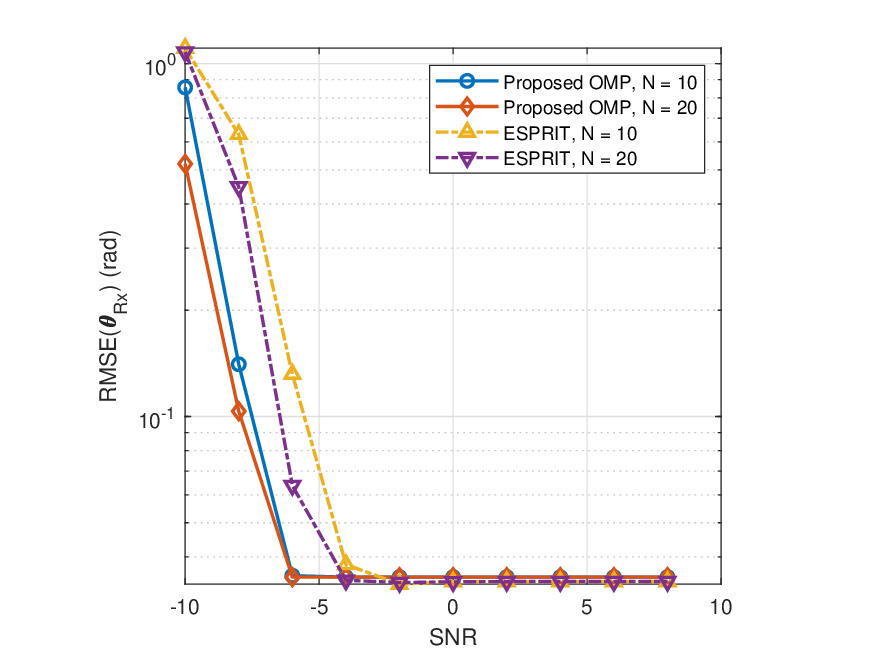}
    \caption{RMSE of AOA versus SNR (dB) in the coarse estimation step.}
    \label{rs2}
\end{figure}
For the RMSE evaluation, we compare the performance of the proposed OMP-based sensing method to the conventional ESPRIT algorithm \cite{b5}. For the ESPRIT algorithm, we construct the matrix $\mathbf{Y}$ by arranging the echo signal over $N$ subcarriers side by side to form $\mathbf{Y} = [\mathbf{y}_s[1], \mathbf{y}_s[2],\ldots, \mathbf{y}_s[N]]$ and use it as the input for the standard multi-snapshot ESPRIT algorithm. Fig.~\ref{rs2} presents the AOA estimation performance evolution with variable SNR of the two methods. It is evident that our proposed method outperforms ESPRIT, especially in the low SNR range (from $-14$ to $-2$ dB). The small gap between the two methods indicates that ESPRIT requires more measurements (snapshots) or a higher SNR than the OMP-based algorithm to achieve the same RMSE level. This could be attributed to the suboptimal nature of the subspace method or an insufficient number of measurements for ESPRIT to construct an accurate covariance matrix. In the same method, the improvement between the case $N = 10$ and $N = 20$ of the OMP-based algorithm is better than that of ESPRIT. In the high SNR range (greater than $-2$ dB), the proposed method and ESPRIT exhibit similar performance, with both algorithms tending to saturate at higher SNR levels.
\subsection{Fine Estimation Performance}
\begin{figure}
    \centering
    \includegraphics[width=1.05\linewidth]{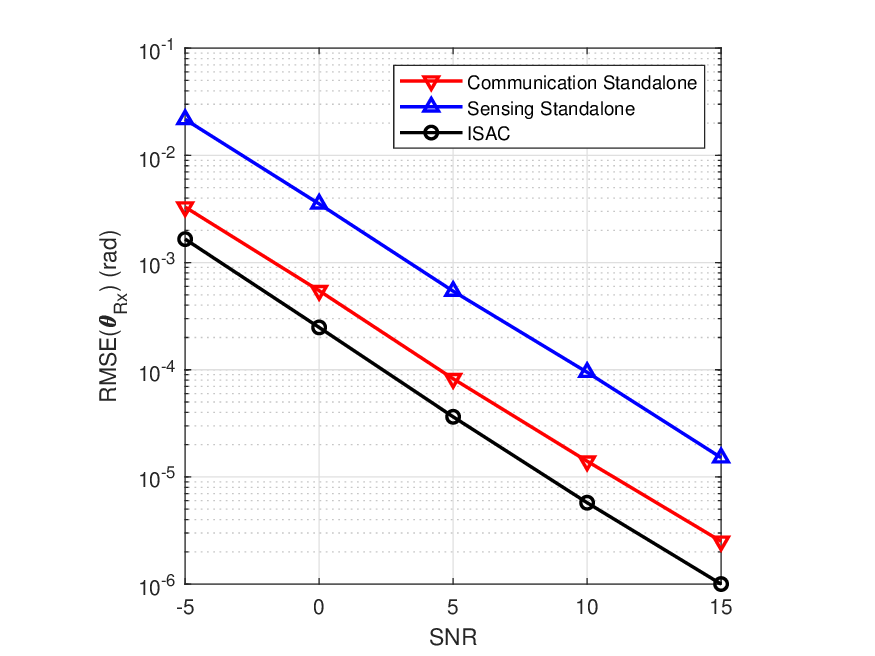}
    \caption{RMSE of AOA vs. SNR (dB) in the refinement step}
    \label{rs3}
\end{figure}

We evaluate Algorithm 2 across three scenarios: an ISAC system with shared AOA, a standalone communication system, and a standalone sensing system. In this evaluation, we focus solely on its performance in terms of the RMSE of AOA, rather than the entire reconstructed channel, which also depends on other parameters such as AOD and path gain. The simulation results are presented in Fig. \ref{rs3}. Notably, the RMSE in the AOA-shared case is lower than in the non-AOA-shared case at the same SNR level. This improvement arises from leveraging information from both the communication and sensing subsystems. We conclude that the joint case, with access to more information, achieves a better AOA RMSE. Specifically, the joint estimation reduces the RMSE by a factor of $2.5$ compared to the standalone communication system and nearly $15$ times compared to the standalone sensing system.
\section{Conclusion}
This paper presents a novel approach to channel estimation in mmWave scenarios by leveraging sensing information from the ISAC system. To improve the accuracy of both sensing and communication functions, we formulated ISAC processing as a sparse signal recovery problem and proposed a novel OMP-based method for sensing parameter estimation. Additionally, we introduced an estimator and analysis demonstrating that sensing and communication functions can mutually enhance overall channel estimation performance, particularly when the sensing target serves as a scatter point. In the future, we will conduct a comprehensive investigation into the interdependencies between subsystems and the interactions between sensing and communication paths within the sensing and communication channels to develop a more effective estimator.

\end{document}